\documentclass{ws-p8-50x6-00}

\begin{document}

\title{The gluon plasma at RHIC}

\author{D\'enes Moln\'ar}

\address{Physics Department, Columbia University\\ 538 West 120th Street, New York, NY 10027, U.S.A. \\
E-mail: molnard@phys.columbia.edu}


\maketitle

\abstracts{
Differential elliptic flow 
and particle spectra are calculated from covariant Boltzmann transport theory
taking into 
account the finite transport opacity of the gluon plasma produced in Au+Au at
$E_{cm}\sim 130$ $A$ GeV 
at RHIC.
The solutions are shown to depend mainly
on the transport opacity, $\chi=\int dz \sigma_t\rho_g$.
The elliptic flow saturation pattern reported by STAR
indicates that $\chi_{b=0} \sim 25$,
i.e., the gluon plasma is $\sim 80$ times more opaque than the pQCD estimate
based on HIJING.
Such large opacities are also consistent with the measured charged hadron spectra.
}

\section{Introduction}
\label{Section:intro}

\vspace*{-0.1cm}
At present we have a very limited understanding of 
the properties of the partonic
environment created in heavy-ion collisions at RHIC.
Predictions for the density of the gluons produced
vary by a factor of five depending on the model considered.
Elliptic flow, $v_2(p_\perp)=\langle \cos(2\phi)\rangle_{p_\perp}$,
 the differential second moment of the azimuthal momentum distribution,
and the high-$p_\perp$ suppression of the particle spectra
have been the subject of increasing interest\cite{hydro,Zhang:1999rs,molnar_v2,gvw,v2_cascade}
because they provide important constraints on the 
density and effective energy loss of partons.

The origin of the remarkable saturation of elliptic flow
$v_2(p_\perp) \to 0.2$ above $p_\perp \sim 2$ GeV reported by STAR at Quark
Matter 2001\cite{STARv2} is an open question for theory.
Calculations
based on inelastic parton energy loss\cite{gvw}
do predict saturation or decreasing $v_2$ at high $p_\perp$.
These calculations are valid for high $p_\perp$,
where collective transverse flow from lower-$p_\perp$ partons can be neglected
and Eikonal dynamics is applicable.
However, a constant spatial anisotropy was assumed
throughout the evolution,
while in reality, it decreases and probably even changes sign.
This is likely to reduce the generated elliptic flow much below\cite{gvw} the
preliminary data.

Ideal hydrodynamics\cite{hydro}, the simplest theoretical framework to study elliptic flow,
agrees remarkably well with the measured elliptic flow data%
\cite{STARv2}
up to transverse momenta $\sim 1.5$
GeV$/c$.
However,
it fails to saturate at high $p_\perp>2$ GeV as does the preliminary
data.

A theoretical problem with ideal hydrodynamics is that it
assumes local equilibrium throughout the whole evolution.
This idealization is marginal
for conditions encountered in heavy ion collisions\cite{nonequil}.
Covariant  Boltzmann transport theory provides a convenient framework
for nonequilibrium dynamics
that depends on the local mean free path $\lambda(x) \equiv 1/\sigma n(x)$.

Parton cascade simulations\cite{Zhang:1999rs,molnar_v2}
show on the other hand, that  the initial parton density
based on HIJING\cite{Gyulassy:1994ew}
is too low  to produce the observed  elliptic flow
unless the pQCD cross sections are artificially enhanced 
by a factor $\sim 2-3$.
However,
gluon saturation models\cite{Eskola:2000fc}
predict up to five times higher initial densities,
and these  may be dense enough to generate the observed collective flow
even with pQCD elastic cross sections.

In this study,
we explore the dependence of elliptic flow and the high-$p_\perp$ suppression
of the particle spectra on the initial density
and the elastic $gg$ cross section.
Though parton cascades
lack at present covariant
inelastic energy loss, 
elastic energy loss alone may  account
for the observed high-$p_\perp$ azimuthal flow pattern
as long as the number of elastic collisions is 
large enough\cite{molnar_v2}.

\section{Covariant parton transport theory}
\label{Section:transport_theory}

\vspace*{-0.1cm}
We consider here, as in \cite{molnar_v2,nonequil,Yang,Zhang:1998ej},
the simplest nonlinear form of Lorentz-covariant Boltzmann transport theory
in which the on-shell phase space density,
evolves with an elastic $2\to 2$ rate.
We solve the transport equation via the MPC algorithm\cite{nonequil},
which maintains Lorentz covariance using the parton subdivision
technique\cite{Yang,Zhang:1998tj}.
See Ref. \cite{molnar_v2} and references therein for details.

For a given nuclear geometry and formation time,
the solutions of the nonlinear transport equation has been shown\cite{molnar_v2}
to depend mainly on the {\em transport opacity}
$\chi\equiv\int dz \sigma_t\rho_g
=N\langle\sin^2 \theta_{cm}\rangle$
and the impact parameter $b$.
Here $\sigma_t$ is the elastic {\em transport} cross section, $N$ is the average number of collisions per parton during the whole evolution, 
while $\theta_{cm}$ is the collision deflection angle in the c.m. frame. 
To good accuracy the transport opacity
factorizes\cite{molnar_v2} as $\chi = C(b) \sigma_t(T_0) dN_g/d\eta$.

We label our results by the transport opacity $\chi$ and
impact parameter $b$.
Furthermore, we quote our transport opacities {\em relative} to that for
the pQCD minijet gluons predicted by HIJING [$dN/d\eta = 210$, $\sigma_{gg\to gg} \approx 3$ mb $\Rightarrow \sigma_t \approx 1$ mb, $\chi_{b=0} \approx 0.3$].

\vspace*{-0cm}
\begin{figure}[h]
\center
\leavevmode
\hbox{
    \epsfysize 4cm
    \epsfbox{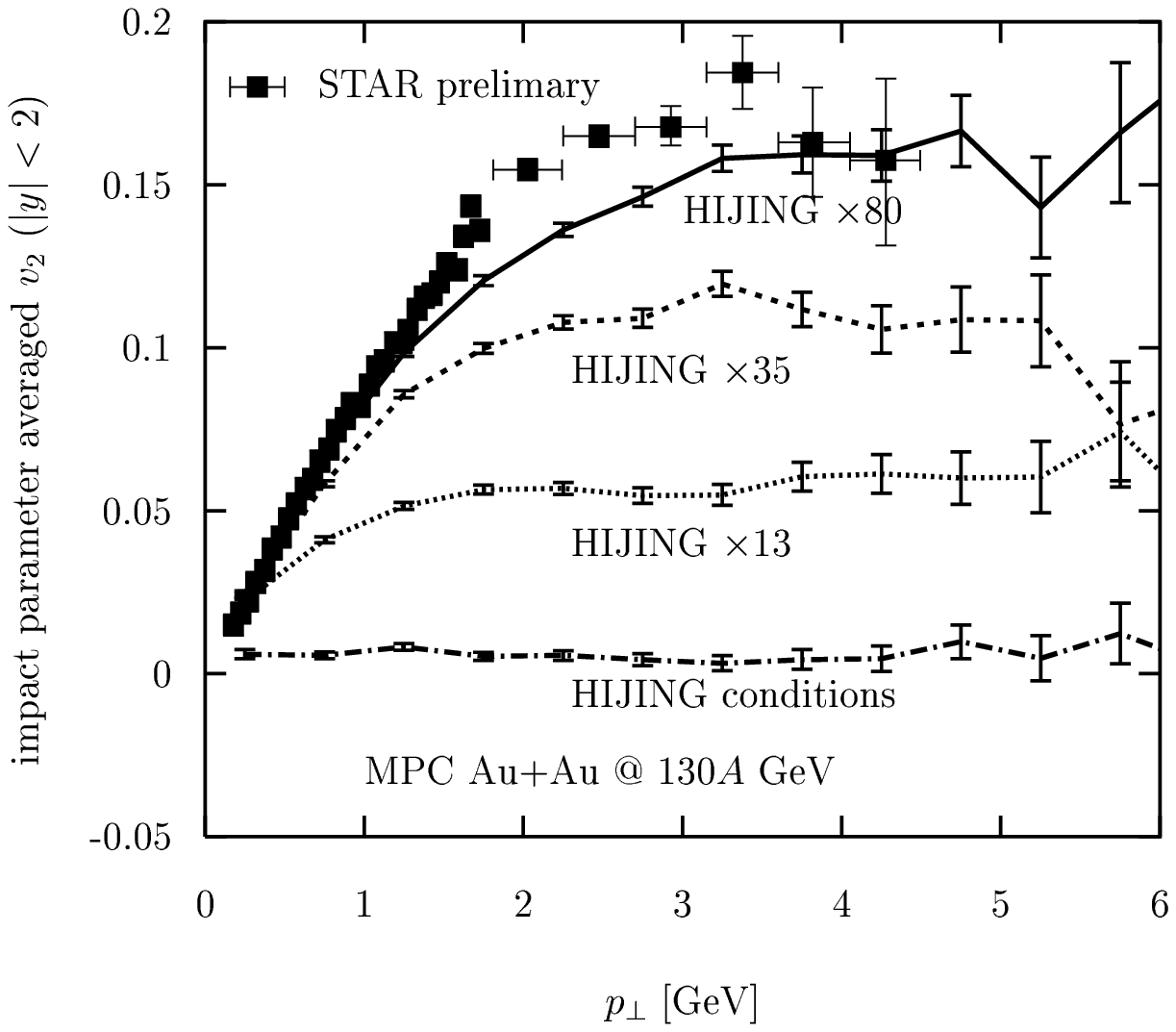}
\hskip 0.5cm
    \epsfysize 4cm
    \epsfbox{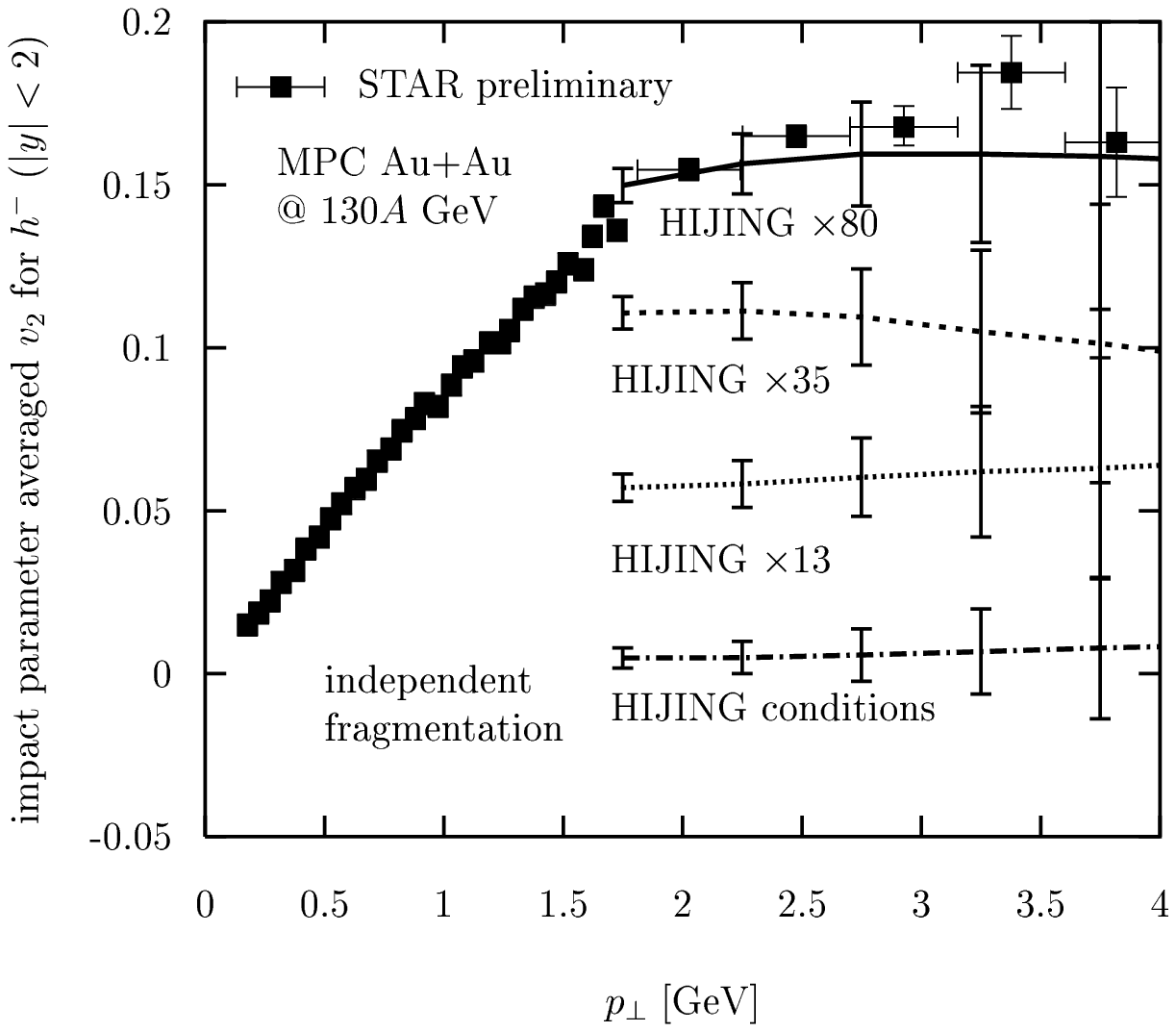}
}
\vspace*{-0.4cm}
\caption{
\footnotesize
Impact-parameter-averaged elliptic flow as a function of $p_\perp$ for Au+Au
at $\sqrt{s}=130A$ GeV for different transport opacities
with hadronization via local parton-hadron duality (left)
and
independent fragmentation (right).
\vspace*{-0.4cm}
}
\label{Figure:1}
\label{Figure:v2}
\end{figure}


\begin{figure}[h]
\center
\leavevmode
\vspace*{-0.6cm}
\hbox{
\hskip -0.1cm
    \epsfysize 4cm
    \epsfbox{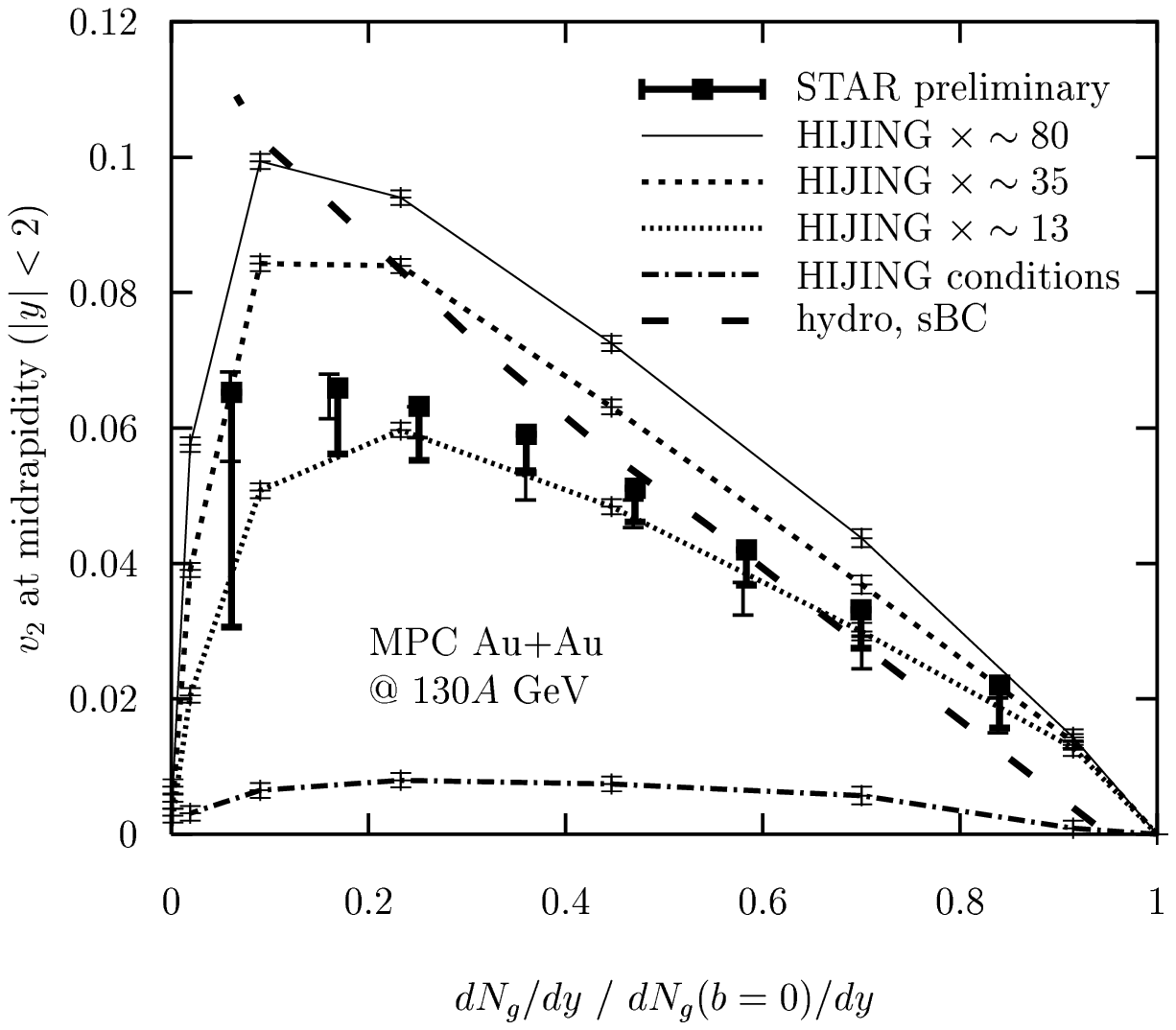}
    \epsfysize 4.5cm
    \epsfbox{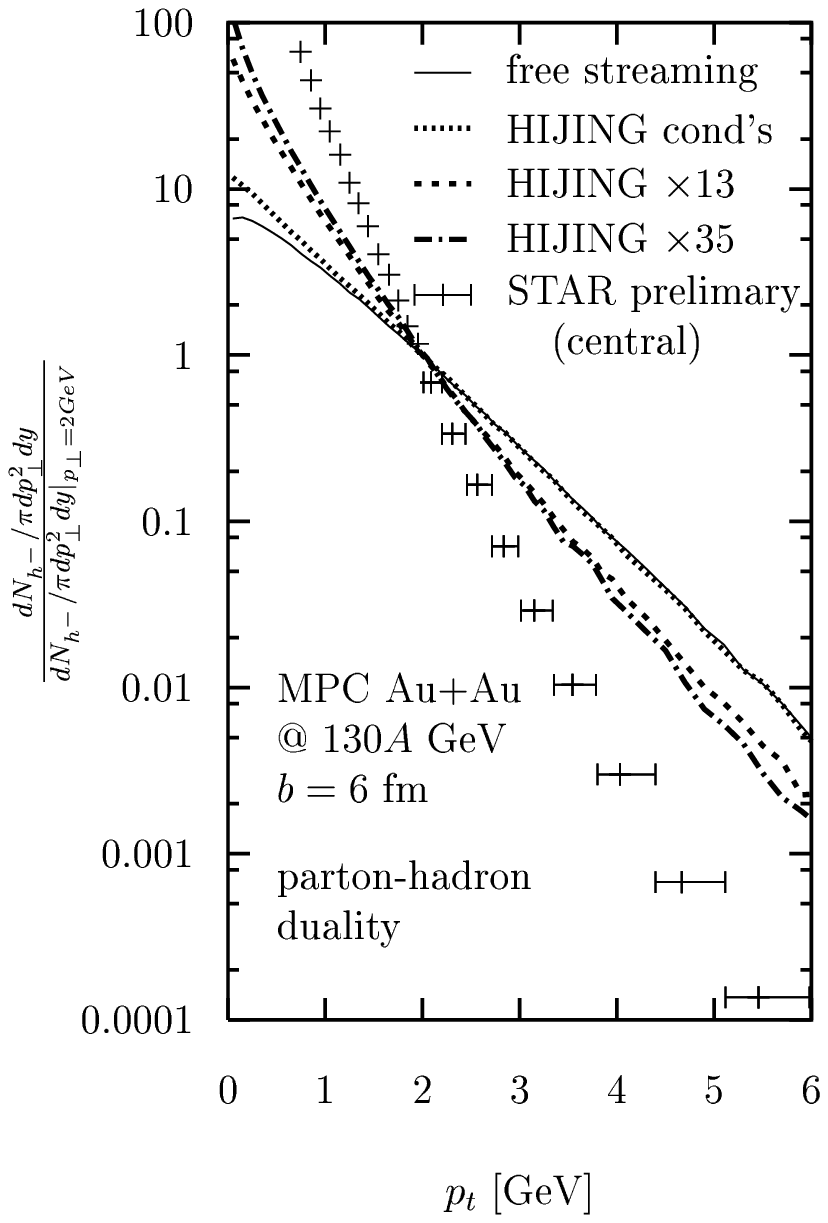}
    \hspace*{-0.5cm}
    \epsfysize 4.5cm
    \epsfbox{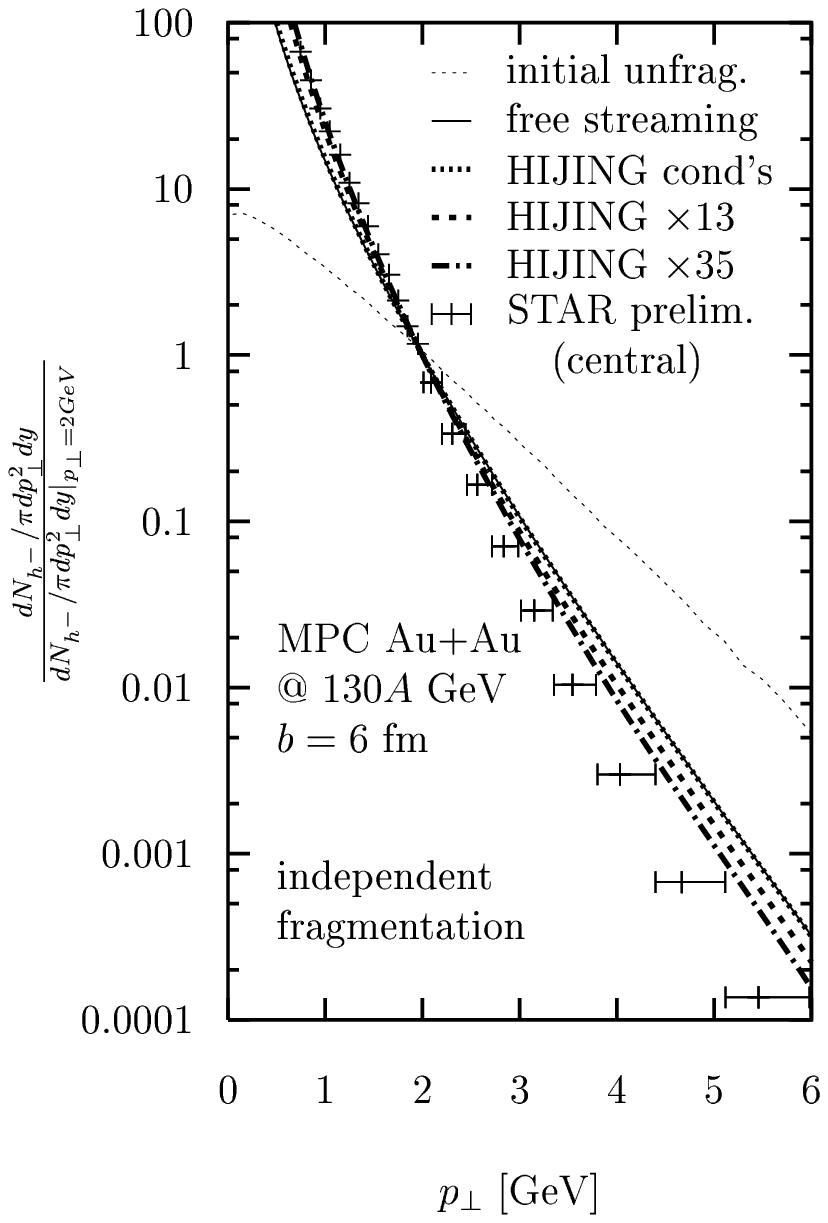}
}
\vspace*{0.1cm}
\caption{
\footnotesize
Charged hadron elliptic flow vs. centrality (left) and negative hadron
$p_\perp$ spectra for $b=6$ fm (middle and right)
as a function of the transport opacity for Au+Au at $\sqrt{s} = 130A$ GeV.
Left and middle plots are for hadronization via parton-hadron duality,
while right plot is for independent fragmentation.
The spectra are normalized to 1 at $p_\perp = 2$ GeV.
\vspace*{-0.7cm}
}
\label{Figure:2}
\label{Figure:pt}
\end{figure}

\section{Numerical results}
\label{Section:glue_results}

\vspace*{-0.1cm}
The initial condition was a longitudinally boost invariant thermal Bjorken tube
at proper time $\tau_0=0.1$~fm/$c$
with uniform pseudo-rapidity distribution between $|\eta| < 5$
and transverse density distribution proportional
to the binary collision distribution for the two gold nuclei.
Based on HIJING, the pQCD jet cross section
was normalized to yield $dN_g/d\eta=210$ in central collisions
and the initial temperature was chosen to be $T_0=700$ MeV.

Two different hadronization schemes were applied.
One is based on local parton-hadron duality,
where each gluon is assumed to convert
to a pion\cite{Eskola:2000fc}.
The other hadronization prescription is independent fragmentation,
where we considered only the $g\to \pi^{\pm}$ channel.
See Ref. \cite{molnar_v2} for details.

Fig. \ref{Figure:v2} shows the impact-parameter-averaged elliptic flow
as a function of $p_\perp$. 
With increasing $p_\perp$,
elliptic flow increases until $p_\perp\sim 1.5-2$ GeV,
where it saturates.
To reproduce the preliminary STAR data,
$\sim 80$ times more opaque initial gluon plasma is needed than the pQCD
prediction from HIJING.
Surprisingly, the results show no sensitivity to the applied hadronization
prescription.

Fig. 2 shows that such large transport opacities are also consistent
with the preliminary charged hadron spectra measured by STAR\cite{Dunlop:2001vh}.
Hadronization via parton-hadron duality yields too little suppression at
high $p_\perp$ because it only incorporates quenching due to elastic energy loss.
However, with the additional quenching due to independent fragmentation,
the parton cascade results approach the preliminary STAR data.

\vspace*{-0.2cm}
\section*{Acknowledgements}

\vspace*{-0.1cm}
We acknowledge the Parallel Distributed Systems Facility
at the National Energy Research Scientific Computing Center
for providing computing resources.

This work was supported by the Director, Office of Energy Research,
Division of Nuclear Physics of the Office of High Energy and Nuclear Physics
of the U.S. Department of Energy under contract No. DE-FG-02-93ER-40764.

\end{document}